\documentclass{ws-procs10x7}
\usepackage{balance,epsfig}

\newcolumntype{d}[1]{D{.}{.}{#1}}

\def\Journal#1#2#3#4{{\it #1} {\bf #2}, #3 (#4)}

\makeindex
\begin{document}

\title{MASS AMBIGUITIES IN CASCADE DECAYS}

\author{B. K. Gjelsten}

\address{Laboratory for High Energy Physics, University of Bern, 
CH-3012 Bern, Switzerland}

\author{D. J. Miller}

\address{Department of Physics and Astronomy, University of Glasgow, 
Glasgow G12 8QQ, U.K.}

\author{P. Osland}

\address{Department of Physics and Technology, University of Bergen, 
N-5020 Bergen, Norway}

\author{A. R. Raklev}

\address{Theory Division, Physics Department, CERN, CH-1211 Gen\` eve, 
Switzerland}


\twocolumn[\maketitle\abstract{
We review the use of invariant mass distributions in cascade decays to
measure the masses of New Physics (NP) particles in scenarios where
the final NP cascade particle is invisible. We extend earlier work by
exploring further the problem of multiple solutions for the masses.
}]

\section{Introduction}

Inspired by the precise determination of the dark matter relic density made
possible with data from the WMAP
satellite\cite{Bennett:2003bz,Spergel:2003cb,Spergel:2006hy} New Physics (NP)
models with a weakly interacting massive particle have flourished over the
last years. Such particles, whether they appear in models of supersymmetry,
universal extra dimensions or little Higgs theories, are invisible to the
detectors of collider experiments, and to be viable dark matter candidates,
certainly stable on the scale of the current age of the universe. The
conservation of a NP quantum number, be it R-parity, KK-parity or T-parity,
ensures a cascade decay of all heavier NP particles into the lightest NP
particle. Measuring NP particle masses in such scenarios by reconstructing
mass peaks is then in general no longer possible.

\begin{figure}[t]
\centerline{\psfig{file=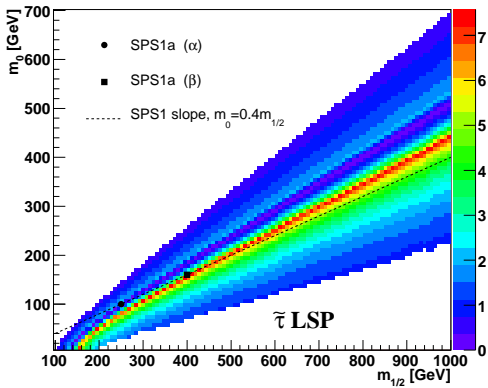,width=2.6in}}
\caption{Scan of $\mu$, as given by Eq.~(\ref{eq:epsilon}), over the SPS1a
mass plane. In the top white area the decay chain (\ref{eq:chain}) is
not kinematically accessible, while the bottom white area has a charged
stau LSP.}
\label{fig:delta_m}
\end{figure}

The well studied alternative to this is to use measurements of the endpoints
of invariant mass
distributions.\cite{Hinchliffe:1996iu,Bachacou:1999zb,Allanach:2000kt,Lester:2001zx,Gjelsten:2004ki,Weiglein:2004hn,Gjelsten:2005aw} As we can
express endpoint positions in terms of the NP masses, we can in principle
solve for the masses if we have at least as many endpoints as unknown
masses. However, there are non-trivial problems with this method. Generally
the expressions for the endpoints are complicated functions, leading to
possible multiple solutions, and strong correlations between the measured
masses.\cite{Gjelsten:2005sv} As an illustration of this problem, we consider
the decay chain
\begin{equation}
\label{eq:chain}
\tilde q_L \to \tilde \chi_2^0 q\to\tilde\ell_R\ell_n q 
\to \tilde \chi_1^0\ell_f\ell_n q
\end{equation}
and show in Fig.~\ref{fig:delta_m} the quantity
\begin{equation}
\mu=\sum_i|m_i^\text{false}-m_i^\text{true}|/m_i^\text{true}
\label{eq:epsilon}
\end{equation}
for a scan over the SPS1a\cite{Allanach:2002nj} mSUGRA
scalar ($m_0$) and gaugino ($m_{1/2}$) mass plane for a
trilinear coupling $A_0=-m_0$, $\tan \beta=10$ and $\mu>0$.
The sum in (\ref{eq:epsilon}) runs over the squark, slepton and
neutralino masses in the decay chain (\ref{eq:chain}).

We find that for virtually all points in the mass plane where the
decay chain (\ref{eq:chain}) exists, and where the lightest
supersymmetric particle (LSP) is not the stau (at low $m_0$), we get a
false solution from the set of endpoints. 
Along the middle of the physical wedge in this plane, and intersecting
the SPS1a line, the above quantity can become rather large.
In an experiment one might also encounter the problem of ``feet'',
i.e. structures in the distributions near the endpoints that are hidden by
background.\cite{Gjelsten:2004ki,Miller:2005zp}

We have previously started an investigation into the possible
improvements to this method that can result from fitting the whole
invariant mass distribution to an analytical expression derived for
its shape.\cite{Miller:2005zp} We found that knowledge of the shape
will predict any possible irregular features at the endpoints for a
measured set of masses, thus essentially removing the problem of
feet. Here we want to investigate further what effect knowledge of the
shape of the invariant mass distribution has on the problem of
multiple solutions.

\section{Multiple solutions}

\begin{figure}[t]
\centerline{\psfig{file=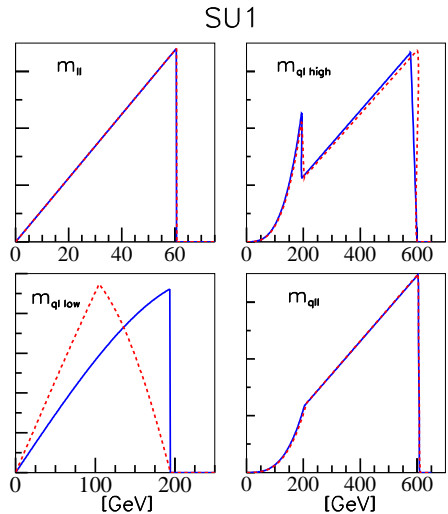,width=3.0in}}
\caption{Invariant mass distributions for the SU1 benchmark
point and the decay chain (\ref{eq:chain}) with arbitrary
normalisation. In blue (solid) are the distributions for the nominal
masses, in red (dashed) are the distributions for the masses in the
false solution.}
\label{fig:su1}
\end{figure}

Using as our example the four possible\footnote{We can form the
invariant mass of the two leptons, the two leptons and a jet, but
assuming that we cannot, in general, identify which lepton is the
lepton nearest to the squark in the decay chain, we use jet-lepton
combinations that give the highest and lowest invariant mass.}
invariant mass distributions for the SUSY decay chain (\ref{eq:chain})
we show in Fig.~\ref{fig:su1} the shape of these distributions, as
found in Ref.~\refcite{Miller:2005zp}, for the nominal masses of the
ATLAS SU1 stau-coannihilation benchmark point. (One may construct other
invariant mass distribution by imposing cuts, but for simplicity we
will neglect them here.)

For SU1 there is also another set of masses that give the same
endpoints for the invariant mass distributions. If only endpoints are
measured we have two indistinguishable solutions for the SUSY
masses. The two sets are given in Table~\ref{tab:masses} as the
nominal masses of the benchmark point, and the false solution from the
nominal endpoints of the benchmark point.

\begin{figure}[b]
\centerline{\psfig{file=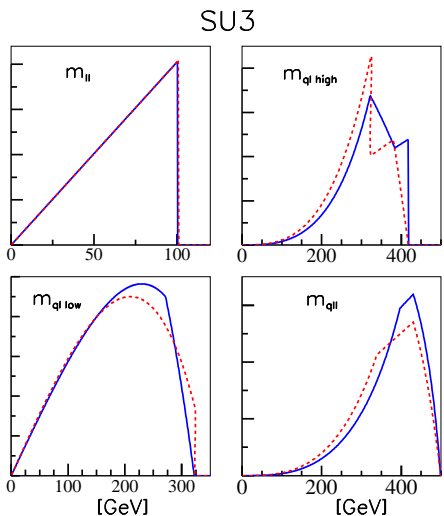,width=3.0in}}
\caption{Invariant mass distributions for the SU3 benchmark
point. See caption of Fig.~\ref{fig:su1} for details.}
\label{fig:su3}
\end{figure}

\begin{table}
\tbl{Nominal and false sets of masses for benchmark points [GeV].
\label{tab:masses}}
{\begin{tabular}{@{}lcccr@{}}
\toprule
Set             & $m_{\tilde\chi_1^0}$ & $m_{\tilde l_R}$ & $m_{\tilde\chi^0_2}$ & $m_{\tilde q_L}$\\ \colrule
SU1 nom.  & 137 & 254 & 264 & 760 \\
SU1 false & 122 & 127 & 246 & 744 \\
SU3 nom.  & 118 & 155 & 219 & 631 \\
SU3 false & 347 & 411 & 452 & 900 \\
\botrule
\end{tabular}}
\end{table}

In Fig.~\ref{fig:su1} we see that there are only small visible
differences in the shapes between the two mass sets of SU1 for three
of the distributions, but for $m_{ql\rm{(low)}}$ there is a marked
change. Similar differences in the shapes can also be seen in three of
the distributions for SU3, the bulk-region benchmark point, shown in
Fig.~\ref{fig:su3}, although they are individually somewhat less
distinct than for SU1. For both cases it seems likely that the false
solution could be ruled out, with the expected high statistics at the
LHC for SUSY scenarios with sub-TeV squark masses. For the SU3
benchmark point the large differences in mass between the solutions
should also show up clearly in the measured cross sections.

This is unfortunately not the situation for all models. For
the SPS1a bulk-region benchmark point\cite{Allanach:2002nj} the
differences in the shapes are very small.  After adding smearing from
detector effects, combinatorics from picking the correct jet, initial
and final state radiation etc., separating the two mass sets for SPS1a
will be very difficult and require very large statistics and an 
extremely good understanding of the detector, if at all possible.

\begin{figure}[b]
\centerline{\psfig{file=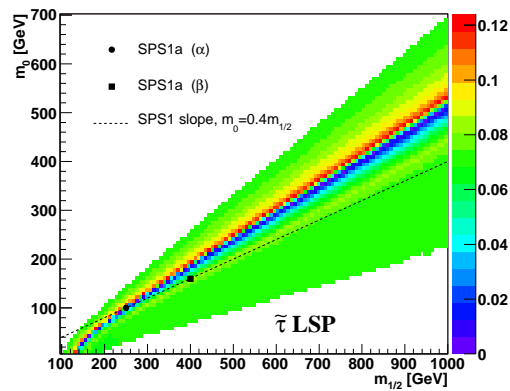,width=2.6in}}
\caption{Scan of $D$, as given by Eq.~(\ref{eq:D}), over the SPS1a
mass plane. In the top white area the decay chain (\ref{eq:chain}) is
not kinematically accessible, while the bottom white area has a charged
stau LSP.}
\label{fig:scan}
\end{figure}

It is then natural to ask how effective the shape can be in distinguishing
multiple solutions over a wide range of SUSY models. In Fig.~\ref{fig:scan} we
show a scan over the SPS1a mass plane. We plot the value of
\begin{equation}
D=\frac{1}{2N}\sum_{i=1}^N\int_0^{m^{\max}_i}|f_i(m)-g_i(m)|\,dm
\label{eq:D}
\end{equation}
where $f_i$ and $g_i$ are the invariant mass distributions for the false and
nominal sets of mass solutions (normalized to $\int dm\, f(m)=1$, $\int dm\,
g(m)=1$), with the $i$-index running over the number of distributions $N$
(here $N=4$), and where $m^{\max}_i$ is the endpoint of the $i$-th
distribution. The numerical pre-factor ensures that $D\leq 1$.

The difference in shape between the nominal and false solutions, as
quantified by the $D$-value is fairly constant at around $D\approx
0.07$, except for a wedge in the mass plane, where we have a
transition between different forms for the analytical expressions for
the $m_{ql\rm{(low)}}$ and $m_{ql\rm{(high)}}$
distributions.\cite{Miller:2005zp} In this region the $D$-value
changes rapidly from low to high values, and from Fig.~\ref{fig:scan}
we see that the SPS1a benchmark point suffers from an unfortunate
position, giving it a value of $D=0.016$. Comparing to
Fig.~\ref{fig:delta_m} we find that while large mass differences can
give large differences in shape, there are certainly areas where the
correct solution could be readily identified from the shape, even when
the mass differences are small, and thus where a comparison of cross
sections would be difficult.

\balance

Due to smearing from the experimental effects mentioned above, it is
difficult to calculate exactly for what $D$-value two sets of
solutions can be distinguished for some given integrated luminosity at
the LHC. However, assuming that the two distributions of SU1 are
distinguishable, and using as a basis for comparison that we find a
$D$-value of $D=0.070$ for SU1, this suggests that we should be able to
distinguish between the nominal and false solutions in most of the
parameter space shown in Fig.~\ref{fig:scan}. We have performed such
scans for different values of $A_0$ and $\tan\beta$, in the mass
planes of the SPS1b, SPS3 and SPS5 benchmark
points,\cite{Allanach:2002nj} and find similar results there.

\section{Conclusions}
We have discussed aspects of the use of analytical expressions for
invariant mass distributions in determining masses in NP scenarios,
using a well studied SUSY decay chain as our example. We have focused
on the possibility of using differences in the shapes of distributions
to remove false solutions that appear if only the endpoints of the
invariant mass distributions are used to determine the NP masses. Our
results from scans of the mSUGRA parameter space indicate that with the
exception of certain limited areas, the shapes of the distributions
should in general be well suited to reject the false solution, given
that enough statistics are available, and may be applied even when a
comparison of cross sections would be difficult.

\section*{Acknowledgments}
ARR acknowledges support from the European Community through a Marie
Curie Fellowship for Early Stage Research Training.  This research has
been supported in part by the Research Council of Norway and the Swiss
National Science Foundation.



\begin{thebibliography}{99}

\bibitem{Bennett:2003bz}
  C.~L.~Bennett {\it et al.},
  \Journal{Astrophys.\ J.\ Suppl.\ }{148}{1}{2003}
  [arXiv:astro-ph/0302207].

\bibitem{Spergel:2003cb}
  D.~N.~Spergel {\it et al.},
  \Journal{Astrophys.\ J.\ Suppl.\ }{148}{175}{2003}
  [arXiv:astro-ph/0302209].

\bibitem{Spergel:2006hy}
  D.~N.~Spergel {\it et al.},
  arXiv:astro-ph/ 0603449.

\bibitem{Hinchliffe:1996iu}
  I.~Hinchliffe, F.~E.~Paige, M.~D.~Shapiro, J.~Soderqvist and W.~Yao,
  \Journal{Phys.\ Rev.\ }{D55}{5520}{1997}
  [arXiv:hep-ph/9610544].

\bibitem{Bachacou:1999zb}
  H.~Bachacou, I.~Hinchliffe and F.~E.~Paige,
  \Journal{Phys.\ Rev.\ }{D62}{015009}{2000}
  [arXiv:hep-ph/9907518].

\bibitem{Allanach:2000kt}
  B.~C.~Allanach, C.~G.~Lester, M.~A.~Parker and B.~R.~Webber,
  \Journal{JHEP}{0009}{004}{2000}
  [arXiv:hep-ph/0007009].

\bibitem{Lester:2001zx}
  C.~G.~Lester,
  CERN-THESIS-2004-003.

\bibitem{Gjelsten:2004ki}
  B.~K.~Gjelsten, D.~J.~Miller and P.~Osland,
  \Journal{JHEP}{0412}{003}{2004}
  [arXiv:hep-ph/ 0410303].

\bibitem{Weiglein:2004hn}
  G.~Weiglein {\it et al.}  [LHC/LC Study Group],
   \Journal{Phys.\ Rept.\ }{426}{47}{2006}
  [arXiv:hep-ph/0410364].

\bibitem{Gjelsten:2005aw}
  B.~K.~Gjelsten, D.~J.~Miller and P.~Osland,
  \Journal{JHEP}{0506}{015}{2005}
  [arXiv:hep-ph/ 0501033].

\bibitem{Gjelsten:2005sv}
  B.~K.~Gjelsten, D.~J.~Miller and P.~Osland,
  arXiv:hep-ph/0507232;
%
  arXiv:hep-ph/0511008.

\bibitem{Allanach:2002nj}
  B.~C.~Allanach {\it et al.},
  in {\it Proc. of the APS/DPF/DPB Summer Study on the Future of
  Particle Physics (Snowmass 2001)}, ed. N.~Graf,
  \Journal{Eur.\ Phys.\ J.\ }{C25}{113}{2002}
  [arXiv:hep-ph/0202233].

\bibitem{Miller:2005zp}
  D.~J.~Miller, P.~Osland and A.~R.~Raklev,
  \Journal{JHEP}{0603}{034}{2006}
  [arXiv:hep-ph/0510356].

\end{thebibliography}
\end{document}